\documentclass{pasj00}
\draft

\Received{yyyy/mm/dd}
\Accepted{yyyy/mm/dd}
\Published{$\langle$publication date$\rangle$}
\SetRunningHead{S.\ Takeuchi et al.}{Modified Slim-Disk Model Based on Radiation-Hydrodynamic Simulation Data}

\usepackage{times,bm}

\begin{document}

\title{Modified Slim-Disk Model Based on Radiation-Hydrodynamic Simulation Data:
\\The Conflict Between Outflow and Photon Trapping}
\author{Shun \textsc{Takeuchi}, 
        Shin \textsc{Mineshige}} 
\affil{%
   Department of Astronomy, Graduate School of Science, Kyoto University, Sakyo-ku, Kyoto 606-8502}
   \and
\author{Ken \textsc{Ohsuga}} 
\affil{%
   National Astronomical Observatory of Japan, Osawa, Mitaka,
Tokyo 181-8588}
\email{E-mail (ST): shun@kusastro.kyoto-u.ac.jp}

\maketitle

\begin{abstract}
Photon trapping and outflow are two key physics associated
with the supercritical accretion flow.
We investigate the conflict between these two processes based on
two-dimensional radiation-hydrodynamic (RHD) simulation data
and construct a simplified (radially) one-dimensional model.
Mass loss due to outflow, which is not considered in the slim-disk model,
will reduce surface density of the flow, and if very significant,
it will totally suppress photon trapping effects.
If the photon trapping is very significant, conversely,
outflow will be suppressed because radiation pressure force will be reduced.
To see what actually occurs, we examine the RHD simulation data
and evaluate the accretion rate and outflow rate as functions of radius.
We find that the former monotonically decreases, while the latter increases,
as the radius decreases. 
However, the former
is kept constant at small radii, inside several Schwarzschild radii,
since the outflow is suppressed by the photon trapping effects.
To understand the conflict between the photon trapping and outflow 
 in a simpler way, we
 model the radial distribution of the accretion rate from the simulation data
and build up a new (radially) one-dimensional model,
which is similar to the slim-disk model but
incorporates the mass loss effects due to the outflow.
We find that
the surface density (and, hence, the optical depth) is much reduced even inside the trapping radius,
compared with the case without outflow,
whereas the effective temperature distribution hardly changes.
That is, the emergent spectra do not sensitively depend on the amount of mass outflow.
We conclude that
the slim-disk approach is valid for interpreting observations,
even if the outflow is taken into account.
The observational implications of our findings are briefly discussed
in relation to ultra-luminous X-ray sources.
\end{abstract}

\section{Introduction}
Supercritical (or super-Eddington) accretion onto black holes remains
one of the most fundamental, classical issues in the present-day
astrophysics
and is now discussed in wide fields of astrophysics
(see Chapter 10 of \cite{Kat+08} for a concise review).
It is well known for the case of spherical accretion
that there is an upper limit to the luminosity; that is
the Eddington luminosity, $L_{\rm E}$.
It is thus impossible for gas to accrete onto a black hole
at a rate exceeding the critical accretion rate,
$\dot M_{\rm crit} \, [\equiv {L_{\rm E}}/(\eta{c^2})]$
where $c$ is the speed of light, $\eta$ is the efficiency.
Then, how about the cases of disk accretion?
Is the supercritical accretion
(accretion at a rate exceeding the Eddington rate) feasible?
This is the enigmatic issue and has been discussed from the 1970's
by many authors, including \citet{ShaSun73},
but it still remains as a controversial issue
because of technical difficulties by the analytical approach.
One of the reasons for these technical difficulties stems from
the multi-dimensional properties of the supercritical accretion flow.

There are two key processes which appear when the disk luminosity
approaches the Eddington luminosity: 
photon trapping and radiation pressure-driven outflow,
and both are multi-dimensional effects.
At very high luminosity, the accretion rate should also be
very high, and so is the optical depth.  
Then, the photon diffusion timescale in the vertical direction
may become shorter than the accretion time of gas.
If this happens, photons generated deep inside the accretion flow
are not able to reach the surface before the material is swallowed by
a black hole.  This is the photon trapping effects
(\cite{Kat77}; \cite{Beg78}; \cite{BegMei82}; \cite{Fla84}; \cite{Blo86}; 
\cite{Col88}; \cite{WanZho99}; \cite{Ohs+02}).
Furthermore, the flow of high luminosities are supported by
radiation pressure, which is likely to induce outflow
(\cite{BisBli77}; \cite{Mei79}; 
\cite{Ick80}; \cite{TajFuk98}).
For complete understanding of the supercritical accretion flow,
we need to solve the multi-dimensional radiation-hydrodynamic (RHD) equations.
This has become possible quite recently thanks to 
the rapid developments of high-speed computers. 

\citet{Ohs+05} were the first to 
succeed in the global RHD simulations of the supercritical accretion flow
until the flow settles down on the quasi-steady phase.
They have demonstrated that the accretion rate can be arbitrarily high,
that the Eddington luminosity can be exceeded,
and that the luminosity increases with an increase of
the mass input rate in a logarithmic fashion
(see also \cite{Ohs06}).
The reason for the occurrence of super-Eddington luminosity
is the combination of two effects: significant radiation anisotropy and
photon trapping (\cite{OhsMin07}).

The slim-disk model was proposed by \citet{Abr+88}
as a simplified model of supercritical accretion flow.
This model is constructed on (radially) the one-dimensional formulation,
like \citet{ShaSun73}, and the photon trapping effects are
expressed as the (radial) advection of radiation entropy,
though the outflow effect is not considered.  Despite this weakness
this model has been used by many authors as a ``standard model"
of the supercritical accretion because of its technical simplicity
(e.g., \cite{Szu+96}; \cite{Bel98}; \cite{WatFuk99}; \cite{Min+00}; \cite{Fuk00}; \cite{Wat+00}; 
\cite{Kaw03}; \cite{GuLu07}).

It is not easy to construct one-dimensional models (like
the slim-disk model) which incorporate both of photon trapping and
outflow effects, since the combined effects are essentially multi-dimensional.
There have been several attempts 
(e.g., \cite{Lip99}; \cite{Kit+02}; \cite{Fuk04}; \cite{Koh+07}; \cite{Pou+07}), but
these studies rely on simplifications and assumptions.
It will be much more preferable to study this issue
by using the multi-dimensional RHD simulations,
but such simulations are expensive and it is not easy to gain physical insight
from simulation data.  In the present study
we thus construct a one-dimensional model as an extension of the slim-disk model
based on the information
regarding the conflict between photon trapping and outflow,
which we gain from the simulation data.
Our goal is to examine how the flow structure and its appearance
are affected by the counteracting effects of photon trapping and outflow.

The plan of this paper is as follows:
In the next section, we overview the RHD simulations of supercritical accretion flow
and calculate the accretion rate and the mass outflow rate of
the simulated flow.  In section 3, we present our improved
slim-disk model and see some details of the flow structure.
We then discuss several important effects of the supercritical
flow based on our simple model.  The final section is devoted to summary.

\section{Supercritical Accretion Flows}
\subsection{Basic Considerations}
Through the studies based on the slim-disk model
two key observational signatures of the supercritical flow have become clear:
smaller innermost radius and the flatter temperature profile
\citep{Wat+00}.
The former is because of large amount of accreting material existing
even inside the radius of the innermost stable circular orbit (ISCO),
emitting significant radiation (see also \cite{WatMin03}).
The latter is due to the suppression of the radiation flux by photon trapping.

The emergent flux distribution of the standard-type disks is
determined by the energy balance, which is approximately expressed by
\begin{equation}
2 \pi R^2 \cdot \sigma T_{\rm eff}^4 \propto 
  \frac{GM {\dot M}_{\rm acc}}{R} \propto R^{-1}.
\end{equation}
This gives the relation, $T_{\rm eff} \propto R^{-3/4}$.
Here, $\sigma$, $G$, $M$, and $ \dot{M}_{\rm acc}$
represent, respectively, the Stefan-Boltzmann constant,
the gravitational constant, mass of the central black hole,
and the mass accretion rate.  

In the slim-disk model, by contrast,
the advection effects, 
the relative importance of which increase inward as
${t_{\rm diff}}/{t_{\rm acc}} \propto R^{-1}$ 
(see Kato et al. 2008, Chap. 10), should be considered.  
As a result, the flux distribution becomes
\begin{equation}
2 \pi R^2 \cdot \sigma T_{\rm eff}^4 \propto 
\frac{GM \dot{M}_{\rm acc}}{R} \cdot \frac{t_{\rm acc}}{t_{\rm diff}}
 \propto R^{0},
\end{equation}
which lead to a somewhat flatter temperature profile,
$T_{\rm eff} \propto R^{-1/2}$.
Here, ${t_{\rm acc}} \, ( = -R/v_R)$ and ${t_{\rm diff}} \, ( = 3H\tau/c)$
represent the accretion timescale and photon diffusion timescale
(in the vertical direction), respectively, and
$v_R$, $H$, and $\tau$ are
radial velocity, scale-height of the disk, and 
optical thickness of the disk, respectively.
The occurrence of photon trapping gives rise to a
critical radius, trapping radius ($R_{\rm trap}$), inside which
photon trapping is significant, and it is expressed as
\begin{equation}
R_{\rm trap} \approx \frac{H}{R}(\dot M_{\rm acc} c^2 / L_{\rm E}) \, r_{\rm s},
\label{eq:Rtrap}
\end{equation}
where $r_{\rm s} \, (= 2GM/c^2)$ is the Schewarzschild radius.

Cautions should be taken to the fact that
the slim-disk model does not perfectly describe
the properties of the supercritical accretion flow, since outflow, 
one of the most important properties of the supercritical flow,
 is not considered.  The mass accretion rate is, hence,
set to be constant in space in the slim-disk formulation.
However, it is likely to decrease inward due to the
mass loss by the outflow.  In the limit of significant mass loss by the outflow,
surface density (and thus the optical depth) of the accretion flow 
will be much reduced, which may result in the total suppression of
the photon trapping effects.  That is, the slim-disk model breaks down.
In the limit of significant photon trapping, conversely, 
(outward) radiation pressure force will be weakened
(or may become even inward, see Ohsuga \& Mineshige 2007), and
the mass loss by the radiation pressure-driven outflow will be negligible.
Which is likely to be the case?  Which process will dominate?

We wish to note that the slim-disk model is not
the only models of the supercritical accretion.
Shakura \& Sunyaev (1973) already discussed the supercritical flow
and considered standard-type disks with large outflow.
They assumed that flux from each radius $F(R)$ cannot exceed the value
that gives the Eddington luminosity, i.e., roughly
$2 \pi R^2 F(R) = L_{\rm E} $
within certain radius, called the spherization radius, $R_{\rm sph}$
(see also \cite{Beg79}).
Using the standard-disk relation [Eq. (1)], we can easily derive
\begin{equation} 
 R_{\rm sph} \approx (\dot M_{\rm acc} c^2 / L_{\rm E}) \,  r_{\rm s}
\label{eq:Rsph}
\end{equation}
Inside this radius, the mass accretion rate decreases inward
in proportion to the radius; i.e., $\dot M_{\rm acc} \propto R$.
This means, the accretion rate vanishes at a very small $R$,
and we again have a somewhat flatter temperature profile,
$T_{\rm eff} \propto R^{-1/2}$,
since $2 \pi R^2 \cdot \sigma T_{\rm eff}^4 \propto \dot{M}_{\rm acc}/R \sim {\rm const.}$

Comparing equations (\ref{eq:Rtrap}) and (\ref{eq:Rsph}),
one can notice that the critical radii for the photon trapping
and outflow are on the same order because $H \sim R$. 
 In other words,
both effects could be equally important in the supercritical flow.
In addition, the apparent effective temperature profiles are same
for both cases.  That is, we cannot simply conclude which effect 
dominates by looking at the observed spectra.

Theoretically speaking, however, there is a big difference.
The photon trapping is more important near the equatorial plane,
since it takes longer time for photons to travel from the equatorial plane,
than from the middle part of the disk, to the disk surface,
while outflow occurs from the disk surface.
This indicates that both effects can be simultaneously significant
but at different heights.
We thus need to carefully examine multi-dimensional simulation data.

\subsection{Model and Simulated Flow}
We simulate the supercritical accretion flow, by using the two-dimensional RHD code 
developed by \citet{Ohs+05}.
The basic equations and the numerical method are described in details in \citet{Ohs+05}.
Hence we briefly summarize them.
We use spherical coordinates $(r,\theta, \varphi)$,
where $r$, $\theta$, and $\varphi$ are the radial distance, the polar angle, 
and the azimuthal angle, respectively, and
set a non-rotating black hole at the origin.
To mimic the general relativistic effects, 
we adopt the pseudo-Newtonian potential,
$\psi$ given by $\psi = -GM/(r-r_{\rm s})$ (\cite{PacWii80}).
As to the flow structure
we assume the axisymmetry  (i.e., $\partial/\partial \varphi = 0$) and 
the reflection symmetry relative to the equatorial plan (with $\theta = \pi/2$).
To solve the radiative transfer,
we apply a flux-limited diffusion (FLD) approximation developed by \citet{LevPom81}.
This approximation is that 
the radiative flux and the radiation pressure tensor are expressed in terms of the radiation energy density
(\cite{TurSto01}; \cite{Ohs+05}).

A difference between the present calculation and \citet{Ohs+05} is 
in the computational domain.
Our purpose in this study is to examine the conflict between outflow and photon trapping.  Therefore,
we have to simulate the supercritical accretion flow in a wider spatial range.
Thus, we set the computational domain spherical shells of 
$3 \, r_{\rm s} \leq r \leq 1000 \, r_{\rm s}$  and 
$0 \leq \theta \leq \pi/2$ and it is divided into $96 \times 96$ grid cells,
 (Note that the computational domain of  \cite{Ohs+05} was $3 \, r_{\rm s} \leq r \leq 500 \, r_{\rm s}$.)
We start the calculations with a hot, rarefied, and optically thin atmosphere. 
There is no cold dense disk initially, 
and we assume steady mass injection into the computational domain through 
the outer disk boundary ($r = 1000\, r_{\rm s}, 0.45\pi \leq \theta \leq 0.5\pi$). 
The injected matter is supposed to have a specific angular momentum corresponding
to the Keplerian angular momentum at $r = 500 \, r_{\rm s}$
(cf. $r = 100 \, r_{\rm s}$ in \cite{Ohs+05}), 
and we set the injected mass-accretion rate (mass input rate) $ \dot M_{\rm input}$ to be constant in time.
Throughout the present simulation, we assume 
$M = 10 M_{\odot}$, $\alpha = 0.1$, $\gamma = 5/3$, $\mu =0.5$, and $Z = Z_{\odot}$.
Here $\alpha$, $\gamma$, $\mu$, and $Z$ are the viscous parameter, the specific heat ratio,
the mean molecular weight, and the metallicity, respectively.

\begin{figure*}
  \begin{center}
  \FigureFile(80mm,55mm){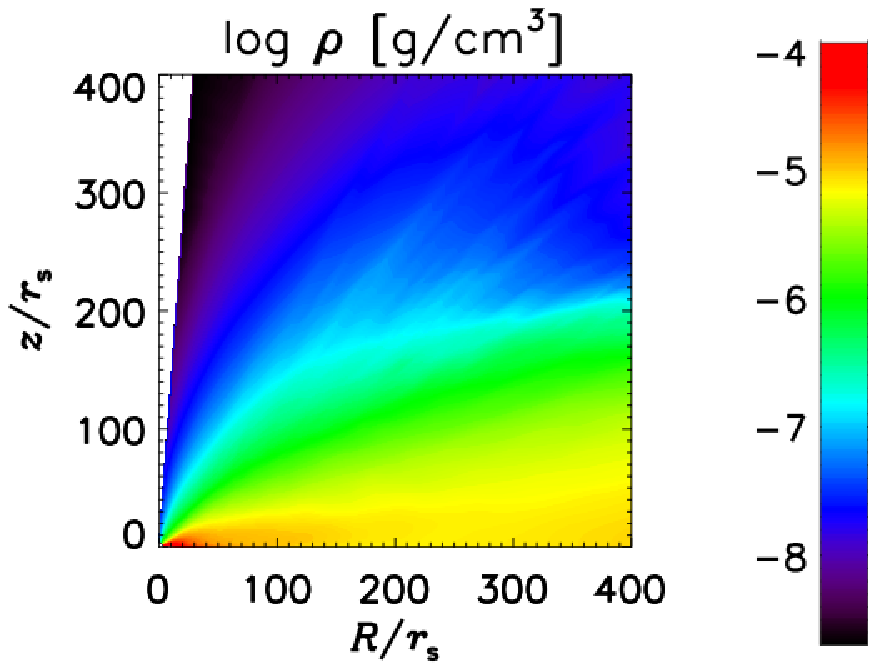}
  \FigureFile(80mm,55mm){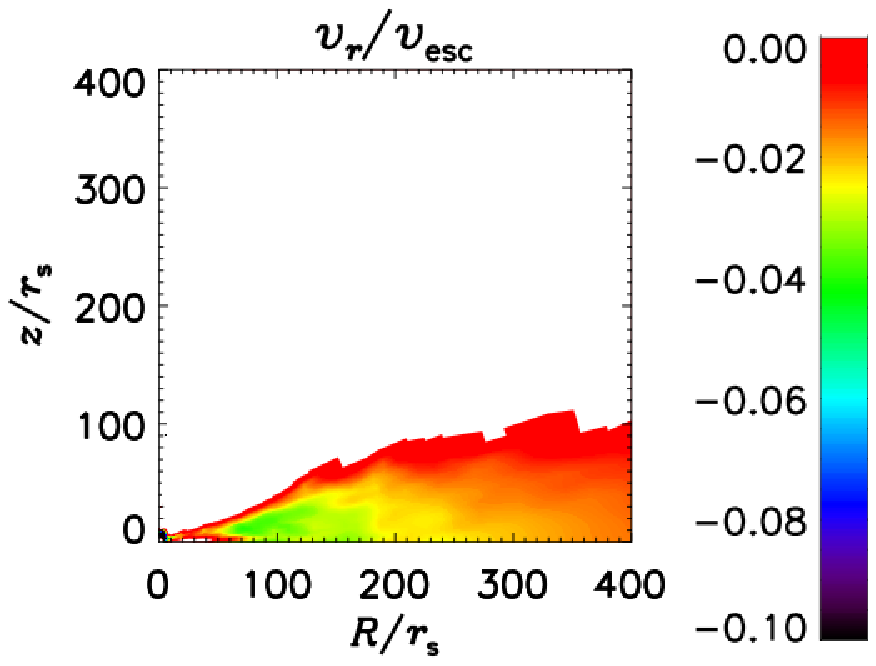}
  \end{center}
\caption{
Color contours of the matter density ({\it left}) and the radial inflow velocity ({\it right}) 
distribution in the meridional section.
Note that these values are time-averaged over $t = 190-250 \, [\rm s]$.
}
\label{fig:simulation}
\end{figure*}

Figure \ref{fig:simulation} indicates the color contours of 
the matter density ({\it left panel}) and the radial inflow velocity ({\it right panel}) 
distribution in the meridional plane, 
which are time-averaged over $t = 190-250 \, [\rm s]$
in the case of $ \dot M_{\rm input} = 1000  \, L_{\rm E}/c^2$.
In the right panel
the radial inflow velocity is normalized by the escape velocity and 
the region with white color indicates the outflow region, 
i.e., $v_r > 0$.
The supercritical accretion flow forms at $r \le 350 \, r_{\rm s}$.
The disk accretion of the high density gas and the outflow of the low density one are clear. 
In the previous calculation, 
the supercritical accretion flow forms at $r \ltsim 80 \, r_{\rm s}$.
The behavior of the present simulated flow in this region is roughly consistent with that of the previous one
(\cite{Ohs+05}; see also \cite{OhsMin07}).

\subsection{Mass Accretion/Outflow Rate}
\begin{figure}
  \begin{center}
  \FigureFile(70mm,50mm){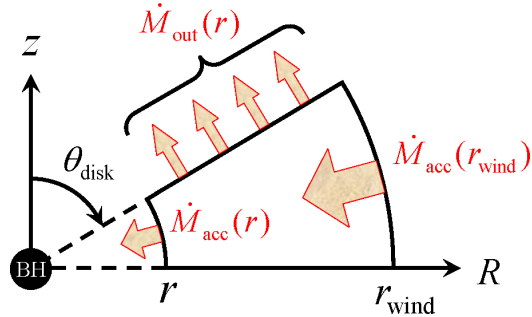}
  \end{center}
\caption{
Schematic picture for the calculation of mass-accretion/outflow rate.
}
\label{fig:schema}
\end{figure}

We assume that an accretion flow occurs in
the region of $\theta_{\rm disk} \le \theta \le \pi/2$,
while outflow region corresponds to 
the region above the accretion flow region, 
and determine an angle $\theta_{\rm disk}$ $(>0)$ from the simulation data.
That is, $\theta_{\rm disk}$ is chosen 
so as to roughly coincide with the angle between the $z$-axis and the boundary 
separating the region with radial inflow (negative velocity) and that of the positive velocity
(figure \ref{fig:schema}).
On the basis of this assumption,
we calculated the mass accretion rate and the cumulative mass outflow rate as functions of radius.
Since the quasi-steady flows form at $r \le 350 \, r_{\rm s}$,
we set in this calculation the outer radius of the flows to be
$r_{\rm wind} = 350 \, r_{\rm s}$.
 
We then calculate the accretion rate, $\dot M_{\rm acc}(r) $,
and the cumulative mass outflow rate, $\dot M_{\rm out}(r) $,
by
\begin{equation}
 \dot{M}_{\rm acc}(r) \equiv \int^{90^{\circ}}_{\theta_{\rm {disk}}}
   4 \pi r^2 \rho v_{r} \sin{\theta}  \, d \theta ,
\end{equation}
and
\begin{equation}
 \dot{M}_{\rm out}(r) \equiv \int^{r_{\rm wind}}_{r}
  4 \pi r' \rho v_{\theta}|_{\theta_{\rm {disk}}}
                      \sin{\theta_{\rm {disk}}} \, d r' .
\end{equation}

\begin{figure*}
  \begin{center}
  \FigureFile(80mm,55mm){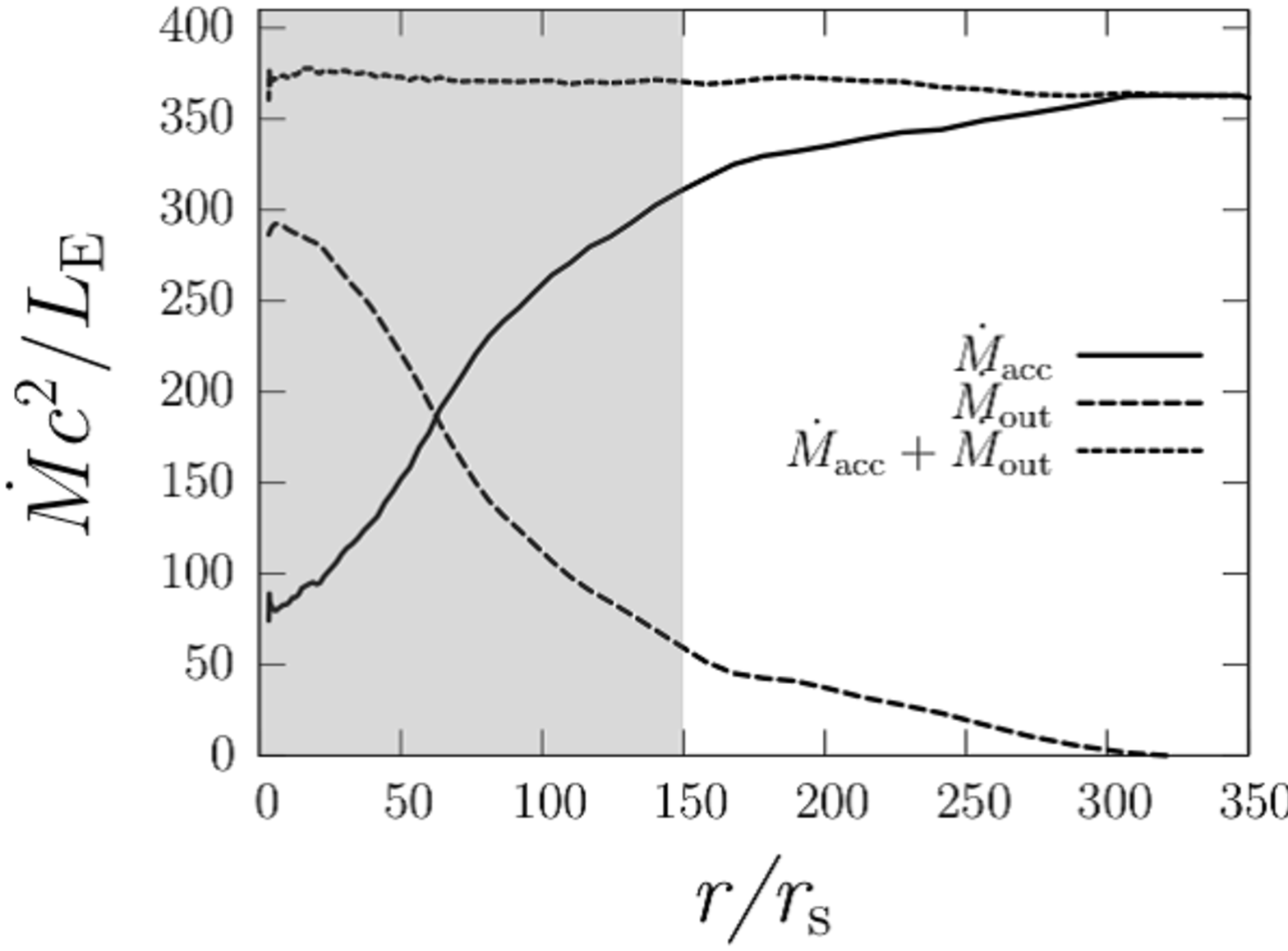}
  \FigureFile(80mm,55mm){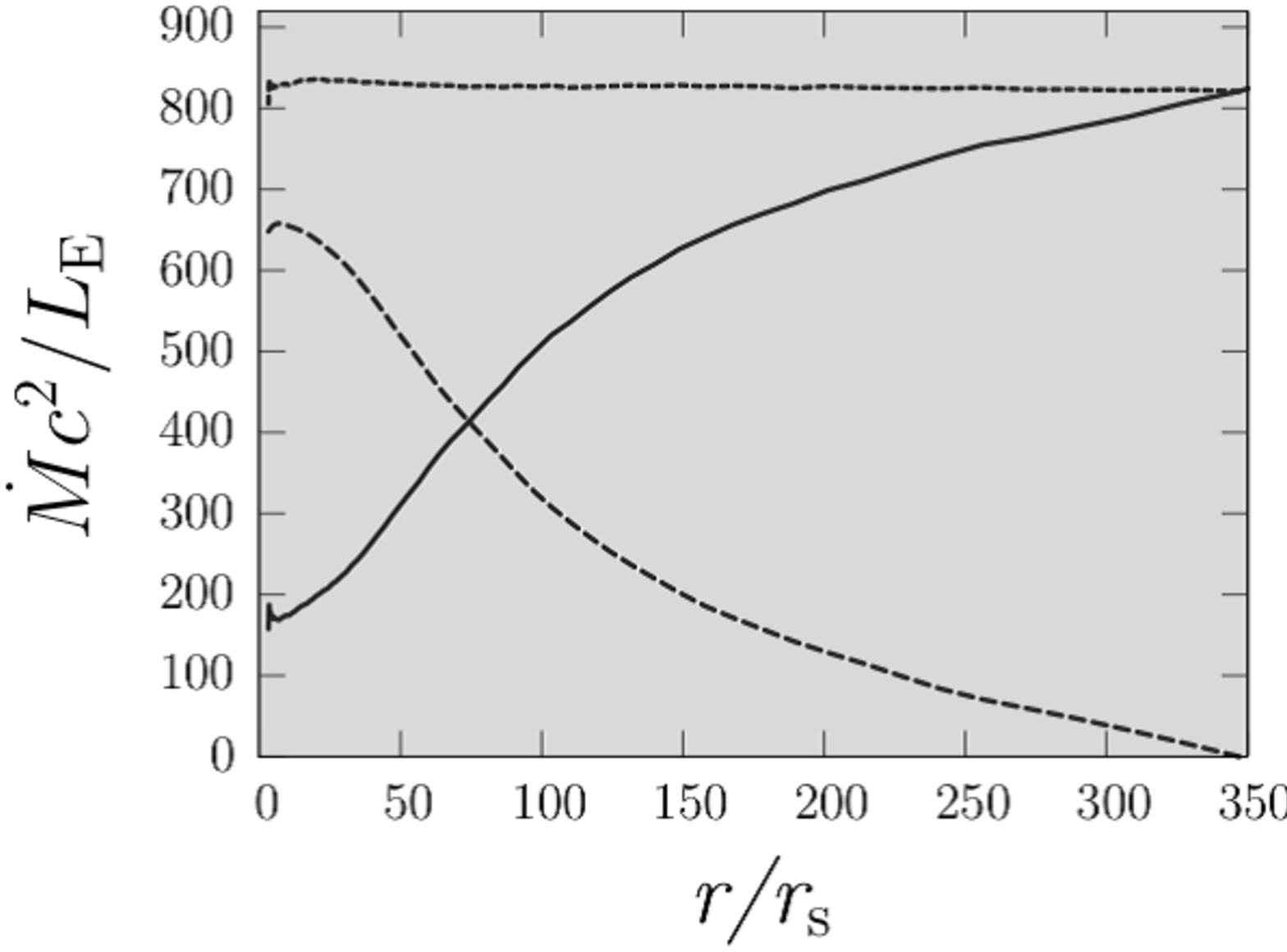}
  \end{center}
\caption{
The accretion rate and cumulative mass outflow rate as functions of radius in the simulated flows
in the case of $ \dot M_{\rm input} = 1000  \, L_{\rm E}/c^2$ ({\it left}) and 
$ \dot M_{\rm input} = 3000  \, L_{\rm E}/c^2$ ({\it right}) .
The solid, dashed, and dotted curves represent the accretion rate  $\dot M_{\rm acc}(r) $, 
mass outflow rate  $\dot M_{\rm out}(r) $ , 
and the sum of the these rates, respectively.
The data of both simulated flows is time-averaged over $t = 190-250 \, [\rm s]$.
The accretion rates at the outer radius $r_{\rm wind}$ ($= 350\, r_{\rm s}$) are 
$\dot{M}_{\rm acc}(r_{\rm wind}) = 364 \, L_{\rm E}/c^2$ ({\it left}) 
and $828 \, L_{\rm E}/c^2$ ({\it right}).
The shadowed areas indicate the photon trapping region.
}
\label{fig:rate}
\end{figure*}

In figure \ref{fig:rate}, 
we show the accretion rate and mass outflow rate as functions of radius 
in the simulated flows for 
$ \dot M_{\rm input} = 1000  \, L_{\rm E}/c^2$ ({\it left panel}) and 
$ \dot M_{\rm input} = 3000  \, L_{\rm E}/c^2$ ({\it right panel}),
respectively.
These values are normalized by the Eddington accretion rate, 
i.e., $L_{\rm E}/c^2$.
Here, the disk inclination angle is chosen to be
$\theta_{\rm disk}=70^{\circ}$ 
 in the case of $\dot M_{\rm input} =1000  \, L_{\rm E}/c^2$, 
and $\theta_{\rm disk}=60^{\circ}$ 
 in the case of $\dot M_{\rm input} =3000  \, L_{\rm E}/c^2$.
In each panel,
the solid, dashed, and dotted curves represent the accretion rate,
$\dot M_{\rm acc}(r) $, 
the cumulative mass outflow rate,  $\dot M_{\rm out}(r) $ , 
and the sum of the these rates, respectively.
In particular,
the following mass conservation should hold,
\begin{equation}
 \dot M_{\rm acc}(r_{\rm wind}) = \dot{M}_{\rm acc}(r) + \dot M_{\rm out}(r) = \rm const.
\end{equation}
We confirm this relation within errors less than $4\%$.
The reason why $\dot M_{\rm acc}(r_{\rm wind}) < \dot M_{\rm input}$ is
because a part of injected gases accumulates at $r_{\rm wind} < r < 1000 \, r_{\rm s}$
or goes out to the outer region of the computational domain without accreting immediately.

The accretion rate is not constant in space due to the mass loss by outflows.
In the analytical accretion flow model, in which an outflow is considered,
the accretion rate decreases inward in proportion to the radius, 
i.e., $\dot M_{\rm acc} \propto r$ (e.g., \cite{ShaSun73}).
That is,  the accretion rate should vanish at a very small radius.  
However, we find that the accretion rates still have a finite value 
at small radii, inside several Schwarzschild radii.
This is because the emergence of outflow is suppressed 
due to the attenuation of radiation flux by photon trapping.
In fact, the radiation flux become even negative (inward) at these radii \citep{OhsMin07}.
The outflows are difficult to blow off 
and thus the flow is easy to be accreted towards the central black hole.

In the case of  $\dot M_{\rm input} =1000  \, L_{\rm E}/c^2$,
the outflow blows off at $r \ltsim 300 \, r_{\rm s}$.
As we evaluated the photon trapping radius by equation (\ref{eq:Rtrap}),
we found that the photon trapping is effective at $r \ltsim 150 \, r_{\rm s}$
for a mass accretion rate of $\dot{M}_{\rm acc}(r_{\rm wind} = 350\, r_{\rm s}) = 364 \, L_{\rm E}/c^2$.
In the case of  $\dot M_{\rm input} =3000  \, L_{\rm E}/c^2$, in contrast,
the outflow and photon trapping is already effective at $r_{\rm wind} = 350 \, r_{\rm s}$.
Thus, we confirmed that the critical radii for the photon trapping and outflow are 
on the same order in the RHD simulations.
Note that this is only a rough estimate, since we assumed
constant accretion rate in the derivation.

\section{Modified Slim-Disk Model}
On the basis of the simulation data analysis presented in the previous section,
we try to construct a modified slim-disk model,
which incorporates the effects of the mass loss by an outflow, in this section.
The accretion flow models which incorporate the mass loss effect were proposed by several researchers
(\cite{Lip99}; \cite{Kit+02}; \cite{Fuk04}; \cite{Pou+07}).
They evaluate the outflow by adopting the spherization radius.
By contrast, we construct a model by using the accretion rate obtained in the previous section.

\subsection{Basic Equations}
We use cylindrical coordinates $(R,\varphi, z)$ in this section.
We assume steady and axisymmetric flow,
and non-rotating black hole,
and adopt a pseudo-Newtonian potential.
We use height-integrated quantities, 
such as $\Sigma = \int^{H}_{-H}\rho \, dz = 2I_{\rm N}\rho H$ and
$\Pi = \int^{H}_{-H}p\,dz  = 2I_{\rm N+1}pH$.
Here, $\Sigma$A$\Pi$A$p$, and $H$ are the surface density , height-integrated pressure,
total pressure, and scale height, respectively.
The coefficient $I_{\rm N}$ and $I_{\rm N+1}$ were introduced by \citet{Hos77}.
The density and the pressure are related to each other by
the polytropic relation, $p \propto \rho^{1+1/{\rm N}}$. 
We assign ${\rm N} = 3$ throughout the entire calculation (i.e., $I_{3} = 16/35$ and $I_{4} = 128/315$).

The continuity equation, the radial component of the momentum equation,
the angular momentum conservation, the hydrostatic equilibrium in the vertical direction,
the energy equation, and equation of state are written as follows:
\begin{equation}
 \dot M_{\rm acc}(R) = \dot{M}_{\rm acc}(R_{\rm wind}) - \dot M_{\rm out}(R) ,  \label{eq:continuity}
\end{equation}
\begin{equation}
 v_R \frac{d v_R}{dR} + \frac{1}{\Sigma} \frac{d\Pi}{dR} 
  = \frac{\ell^2 - \ell^2_{\rm K}}{R^3} 
  - \frac{\Pi}{\Sigma} \frac{d{\rm ln} \Omega_{\rm K}}{dR} , \label{eq:momentum}
\end{equation}
\begin{equation}
\dot{M}_{\rm acc} (\ell - \ell_{\rm in})  = - 2 \pi R^2 T_{R \varphi} ,
\end{equation}
\begin{equation}
(2{\rm N}+3) \frac{\Pi}{\Sigma} = H^2 \Omega_{\rm K}^2  , \label{eq:static}
\end{equation}
\begin{equation}
Q^{+}_{\rm vis} = Q^{-}_{\rm rad} + Q^{-}_{\rm adv}  ,  \label{eq:energy}
\end{equation}
\begin{equation}
\Pi = \Pi_{\rm gas} + \Pi_{\rm rad}
      = \frac{k_{\rm B}}{\bar{\mu} m_{\rm H}} \frac{I_{\rm N+1}}{I_{\rm N}} \Sigma T_{\rm c}  \label{eq:state}
         + \frac{2}{3} I_{\rm N+1} a T_{\rm c}^4 H .
\end{equation}
Here, the flow velocity of radial component and azimuthal one are expressed 
by $v_R$ and $v_{\varphi}$, respectively,
the angular momentum of the gas is given by $\ell$$(=R v_{\varphi}=R^2 \Omega)$,
$\Omega$ and $\Omega_{\rm K}$ are the angular speed of rotation and the Keplerian angular speed,
$\ell_{\rm K}$ and $\ell_{\rm in}$ are Keplar angular momentum and 
the angular momentum at inner radius of the flow,
$T_{R \varphi} (\equiv - \alpha \Pi)$ is vertically integrated stress tensor 
with $\alpha$ being the viscosity parameter (\cite{ShaSun73}),
$\Pi_{\rm gas}$ and $\Pi_{\rm rad}$ are the gas pressure and the radiation pressure,
and $m_{\rm H}$, $T_{\rm c}$, $k_{\rm B}$, and $a$ are 
the hydrogen mass, the temperature on the equatorial plane, the Boltzmann constant,
and the radiation constant, respectively.
The last term on the right-hand side of equation (\ref{eq:momentum})
is a correction term resulting from the fact that the radial component
of the gravitational force changes with height (\cite{Mat+84}).

In the energy equation (\ref{eq:energy}),
the viscous heating rate $Q^{+}_{\rm vis}$, 
the radiative cooling rate $Q^{-}_{\rm rad}$, 
and the advective cooling rate $Q^{-}_{\rm adv}$ are defined by
\begin{equation}
Q^{+}_{\rm vis} 
 = - R \, T_{R \varphi} \frac{d\Omega}{dR} ,
\end{equation}
\begin{equation}
Q^{-}_{\rm rad} 
 = 2 F = \frac{16 a c T^4_{\rm c}}{3 \tau} ,
\end{equation}
\begin{equation}
Q^{-}_{\rm adv} 
= \frac{9}{8}v_R \Sigma T_{\rm c} \frac{ds}{dR} ,
\end{equation}
where $F$ is the radiative flux per unit surface area on the flow surface,
$s$ is the specific entropy,
and $\tau$ is the optical thickness of the flow given by 
$\tau = (\kappa_{\rm es} + \kappa_{\rm ff}) \Sigma$.
$\kappa_{\rm es}(=0.4)$ is the electron scattering opacity,
$\kappa_{\rm ff} (=0.64 \times 10^{23} \bar{\rho} \bar{T}^{-7/2})$ is the free-free absorption opacity,
and $\bar{\rho}$ and $\bar{T}$ are the vertically averaged density and the temperature, respectively.
Note that \citet{Jia+09} point out that the usage of
equation (\ref{eq:static}) may lead to overestimation of gravity
force when the scaleheight is comparable to the radius, $H \sim R$. 

We wish to note that the difference between the equations of  the original slim-disk model and those of our model
is only in the continuity equation (\ref{eq:continuity}).
In the original slim-disk model, the continuity equation is expressed 
by $\dot M_{\rm acc}(R) = \rm {const}$.
The reason why the forms of equations (\ref{eq:momentum})-(\ref{eq:state}) do not change
is because these expressions are written per unit mass. 
While \citet{Lip99} explicitly expresses the outflow effect in her equations,
the physical meanings of the equations are the same as own 
(see also \cite{Kit+02}).
The difference between the previous models and our model is that
we construct a model which realistically considers the outflow effect
by using the accretion rate in the previous section.
In the previous study, moreover,
the flow is approximated to be Keplarian (\cite{Lip99}; see also \cite{Pou+07})
or assumed to obey a self-similar solution (\cite{Kit+02}; see also \cite{Fuk04}),
whereas we solve numerically the radial advection, 
i.e., equation (\ref{eq:momentum}) and (\ref{eq:energy}).
However, we need to take into account the work exerted on the outflow by the accretion flow material.
This issue will be discussed in the next section.

The calculations were performed from the outer radius at $R = 1 \times 10^4 \, r_{\rm s}$
down to the inner radius, $R \sim 1.0  \, r_{\rm s}$.
In our modified slim-disk model,
we approximate
$\dot M_{\rm acc}(R) = \dot M_{\rm acc}(r)$ and
$\dot M_{\rm out}(R) = \dot M_{\rm out}(r)$
within $R = 350 \, r_{\rm s}$.
Since the range of the accretion rate which is obtained by the simulation in the previous section
is at $R \le 350 \, r_{\rm s}$,
we assume that the accretion rate at $R = 350 \, r_{\rm s} - 10^4 \, r_{\rm s}$ is constant, 
i.e., $\dot M_{\rm acc}(R') = \dot{M}_{\rm acc}(R_{\rm wind}) = \dot{M}_{\rm acc}(r_{\rm wind})$
for $350 \, r_{\rm s} \le R' \le 10^4 \, r_{\rm s}$.
In the original slim-disk model, in contrast, 
$\dot M_{\rm acc}(R) = \dot{M}_{\rm acc}(R_{\rm wind})$ in the whole radius.
We set the black hole mass and the viscous parameter 
to be $M = 10 M_{\odot}$ and $\alpha = 0.1$, respectively.
We also set $\dot{M}_{\rm acc}(R_{\rm wind}) = 364 \, L_{\rm E}/c^2$ and $828 \, L_{\rm E}/c^2$.

\subsection{Flow Structure}
\begin{figure*}
  \begin{center}
  \FigureFile(80mm,55mm){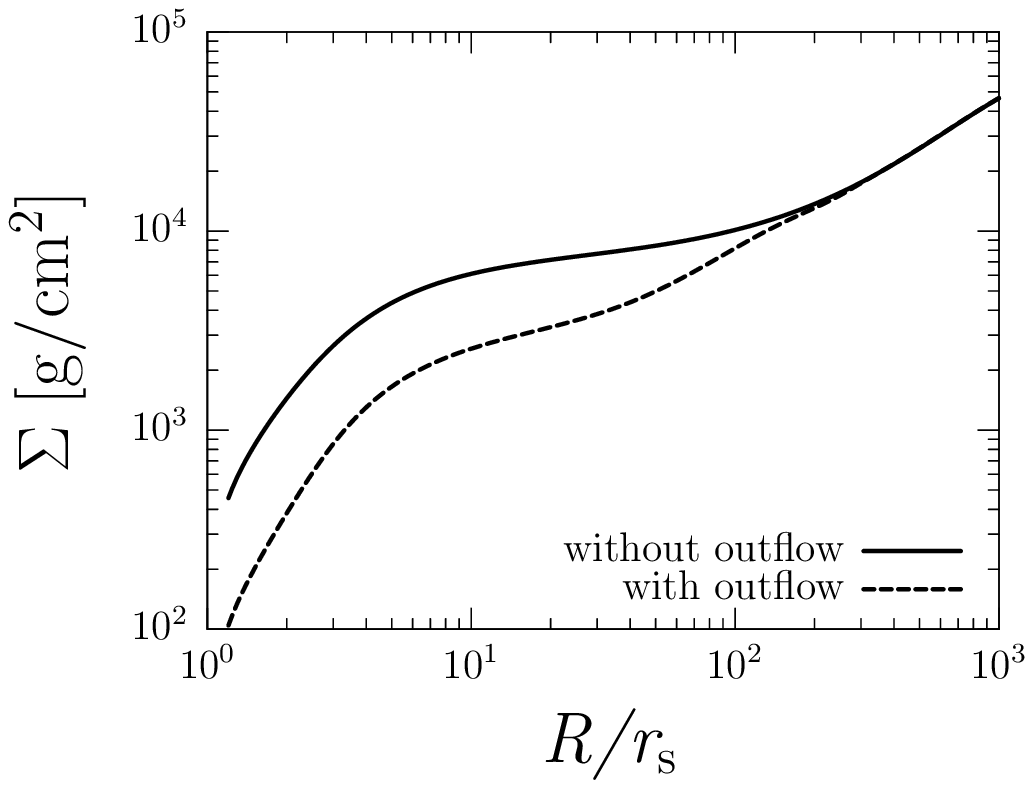}
  \FigureFile(80mm,55mm){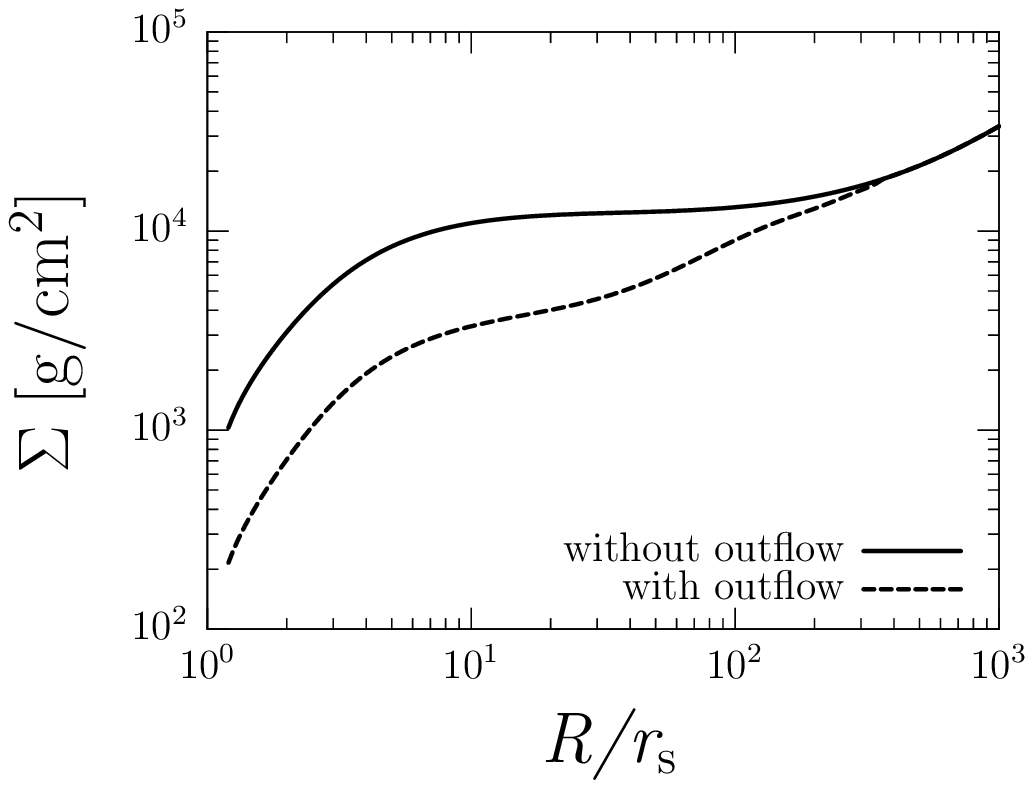}
  \FigureFile(80mm,55mm){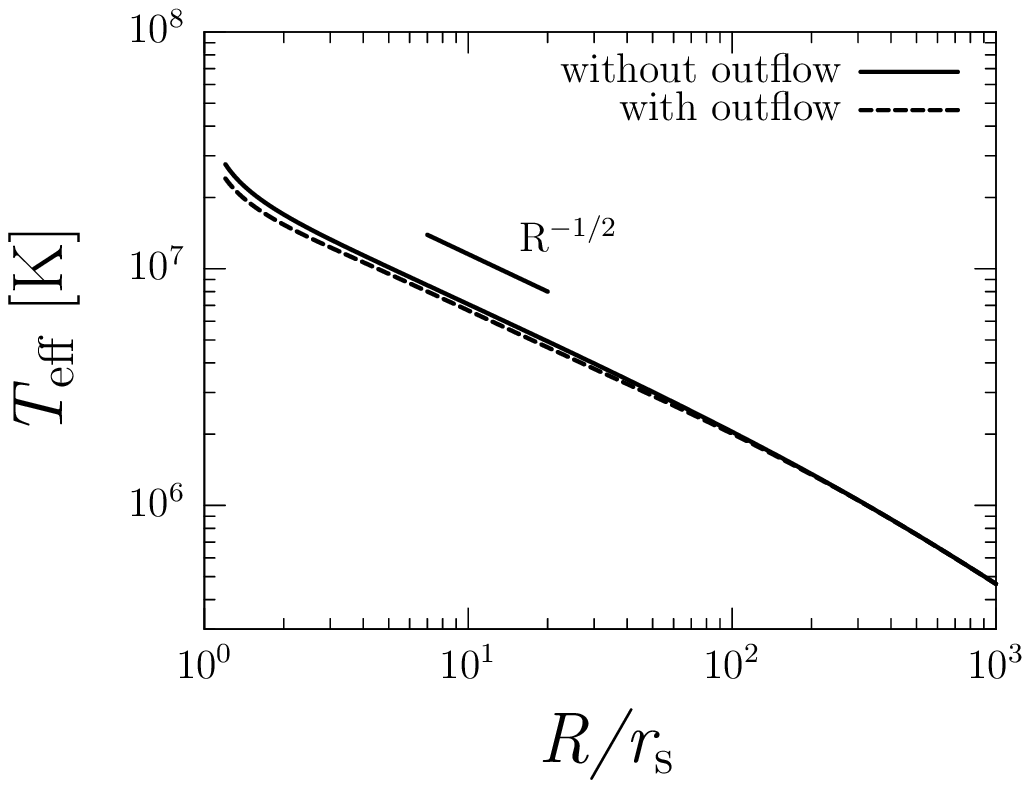}
  \FigureFile(80mm,55mm){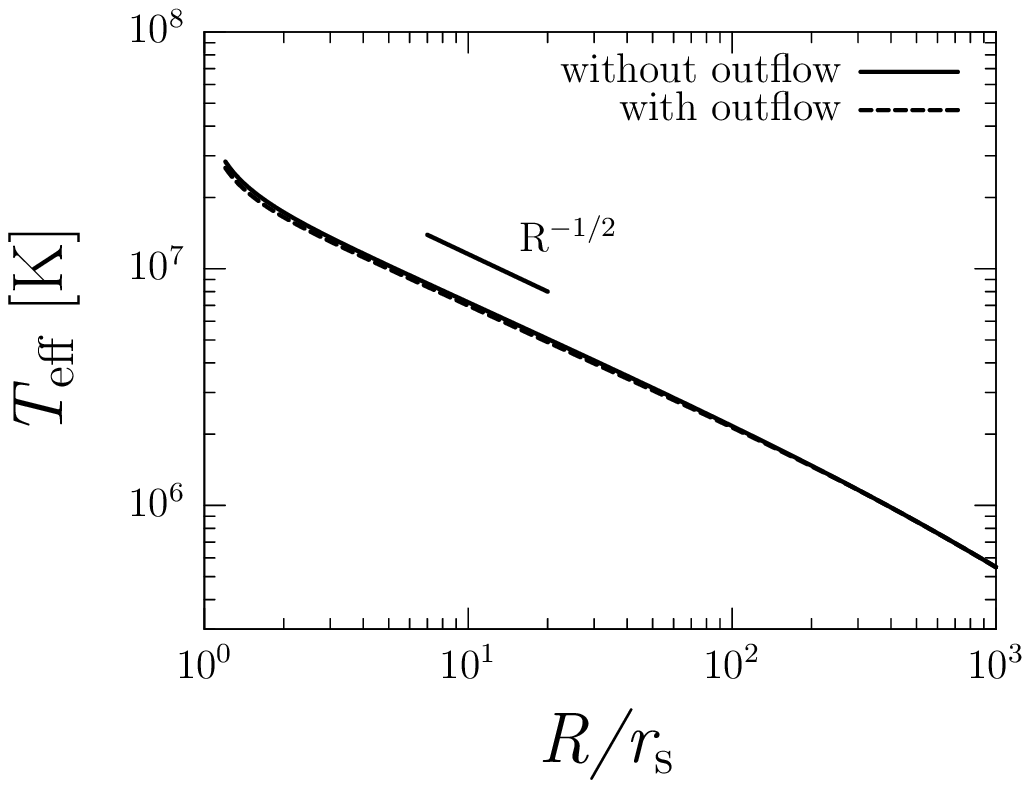}
  \end{center}
\caption{
The surface density $\Sigma$ and the effective temperature $T_{\rm eff}$ as functions of radius
in the modified slim-disk model (dashed lines) and the original slim-disk model (solid lines)
for $ \dot M_{\rm input} = 1000  \, L_{\rm E}/c^2$ ({\it left}) and 
$ \dot M_{\rm input} = 3000  \, L_{\rm E}/c^2$ ({\it right}) .
}
\label{fig:structure}
\end{figure*}

Figure \ref{fig:structure} indicate the surface density and the effective temperature as functions of radius
in the modified slim-disk model (dashed lines) and the original slim-disk model 
(solid lines)
for $ \dot M_{\rm input} = 1000  \, L_{\rm E}/c^2$ ({\it left panels}) and 
$ \dot M_{\rm input} = 3000  \, L_{\rm E}/c^2$ ({\it right panels}), 
respectively.
The surface density of the flow is significantly reduced due to the mass loss 
by outflow in both cases (see figure \ref{fig:rate}).
Hence, the optical depth of the flow $\tau$ $(\propto \Sigma)$ also is reduced
by a factor of two or three.
However, the flow is optically thick in the whole region.
Even though the mass accretion rate decreases as the radius decreases,
the trapping radius derived from the mass accretion rate 
at any radius exceeds that radius; i.e., photon trapping is effective.

We quantitatively examine the effect of photon trapping, i.e., 
we calculate the relative importance of the advective cooling in the energy equation.
 We derive the relation between $t_{\rm diff}/t_{\rm acc}$ and $Q^{-}_{\rm adv}/Q^{+}_{\rm vis}$.
 The advective cooling rate are expressed by
 \begin{equation}
Q^{-}_{\rm adv} 
\sim - \frac{9}{8}\frac{v_R}{R} \Sigma  \left( e - \frac{p}{\rho} \right) 
= - \frac{3}{2}\frac{v_R}{R} H a T_{\rm c}^4,
\end{equation}
where $e$ is the specific energy.
Here, we used $e = aT_{\rm c}^4/\rho$ and $p = aT_{\rm c}^4/3$.
Therefore, the condition of effective photon trapping ($t_{\rm diff}/t_{\rm acc} > 1$) 
is derived by
\begin{equation}
\frac{t_{\rm diff}}{t_{\rm acc}}
= \frac{32}{3} \frac{Q^{-}_{\rm adv}}{Q^{-}_{\rm rad}}
= \frac{32}{3} \frac{Q^{-}_{\rm adv}}{Q^{+}_{\rm vis} - Q^{-}_{\rm adv}} > 1.
\end{equation}
As a result, this gives the relation, 
\begin{equation}
\frac{Q^{-}_{\rm adv}}{Q^{+}_{\rm vis}} > \frac{3}{35} \sim 0.09.
\end{equation}
That is, photon trapping is effective at $Q^{-}_{\rm adv}/Q^{+}_{\rm vis} \gtsim 0.09$.

\begin{figure*}
  \begin{center}
  \FigureFile(80mm,55mm){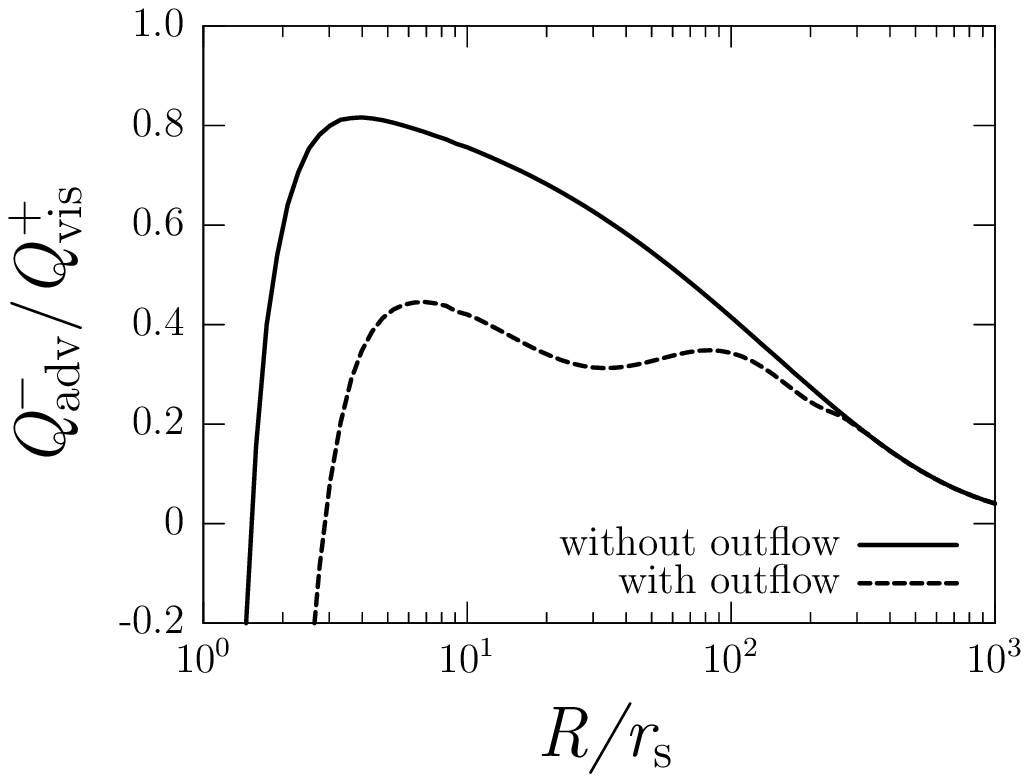}
  \FigureFile(80mm,55mm){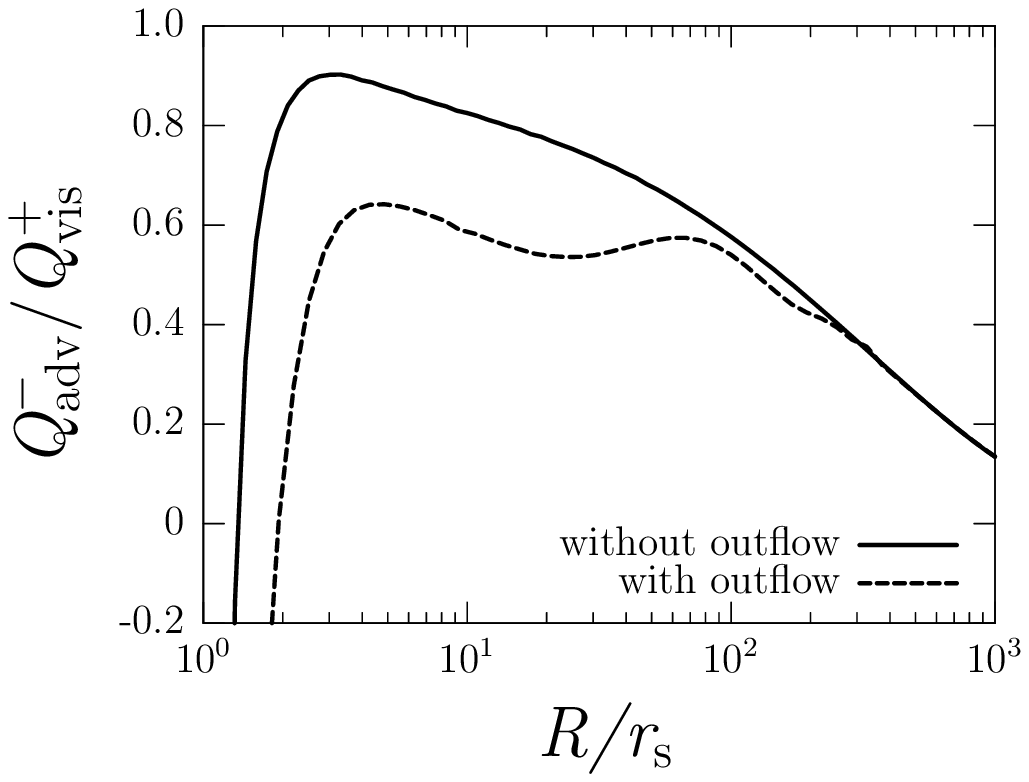}
  \end{center}
\caption{
Ratio of the advective cooling rate $Q^{-}_{\rm adv}$ to the viscous heating rate $Q^{+}_{\rm vis}$ as functions of radius
in the modified slim-disk model (dashed lines) and the original slim-disk model (solid lines)
for $ \dot M_{\rm input} = 1000  \, L_{\rm E}/c^2$ ({\it left}) and 
$ \dot M_{\rm input} = 3000  \, L_{\rm E}/c^2$ ({\it right}) .
}
\label{fig:trapp}
\end{figure*}

In figure \ref{fig:trapp}, we show the ratio,  $Q^{-}_{\rm adv}/Q^{+}_{\rm vis}$, for each model.
In the case of the original slim-disk model, in which the outflow is not considered,
photon trapping is significantly effective at $R \ltsim 100 \, r_{\rm s}$.
In the case of the modified slim-disk model, however,
the advective cooling rate is smaller due to the mass loss by outflow ($Q^{-}_{\rm adv} \propto \Sigma$),
and so is the ratio.
Hence, the photon trapping effect is weaker in the model,
but the outflow is not strong enough to totally suppress photon trapping at the smaller radii 
($Q^{-}_{\rm adv}/Q^{+}_{\rm vis} \sim 0.4-0.6$).
At the larger radius, there is no difference between the two models because there is no outflow.

In addition, 
we understand from figure \ref{fig:structure}
that, even if outflow effects are taken into account,
the effective temperature profiles do not change.
It is $T_{\rm eff} \propto r^{-1/2}$ at the smaller radius, 
and $T_{\rm eff} \propto r^{-3/4}$ at the larger radius. 
Why is then the effective temperature profile unchanged?
This is because it is determined by the photon trapping 
at small radii 
(note that photon trapping is effective at $Q^{-}_{\rm adv}/Q^{+}_{\rm vis} \gtsim 0.09$),
while no significant outflow occurs at large radii .

\citet{Kit+02} construct a model for supercritical accretion flows with mass loss,
adopting the self-similar treatment proposed by \citet{NarYi94}.
They mention that the effective temperature profile of the flow is negligibly affected by occurrence of outflow
in the case of advective-dominated flows (see also \cite{Fuk04}).
The reason is as follows.
When the radiative flux $F$ ($= \sigma T_{\rm c} \propto \Pi$) is reduced by the mass loss,
optical depth $\tau$ ($\propto \Sigma$) is reduced at the same time.
As a result, the effective temperature of the flow $T_{\rm eff}$ $(=T_{\rm c}\tau^{-1/4})$ 
does not depend on the mass loss.
This provides another explanation for the flatter temperature distribution.

In the region at $R \ltsim 3\, r_{\rm s}$,  the advective cooling rate becomes negative.
The reason is as follows.
The optical depth of the flow $\tau$ $(\propto \Sigma)$ steeply decreases inward in that region,
although the flow is optically thick (see figure \ref{fig:structure}).
Therefore, the radiative cooling rate $Q^{-}_{\rm rad}$ $(\propto \tau^{-1})$ steeply increases inward,
whereas the viscous heating rate ($Q^{+}_{\rm vis} \propto \dot{M}_{\rm acc}$) does not.
As a result, the advective cooling rate should become negative,
since  $Q^{-}_{\rm adv} = Q^{+}_{\rm vis} - Q^{-}_{\rm rad}$.

\section{Discussion}

\subsection{Brief summary}
In this paper, we first carefully examined the global RHD simulation data of 
supercritical accretion flow onto black holes
in order to examine the conflict between photon trapping and outflow.  
We have confirmed that both are equally important; i.e.,
despite significant mass loss by the outflow, 
it is not strong enough to totally suppress photon  trapping.
We evaluated the accretion and outflow rates as functions of radius
based on the simulation data
and put them into the formulation of the slim-disk model, in which
no outflows were considered originally and hence
the mass accretion rate was considered to be constant.
We compared the resultant flow structure with consideration of outflow by 
those without consideration of outflow,
finding that, although surface density (and, hence, optical depth) of the
flow is significantly reduced,
the effective temperature profile is negligibly affected by the occurrence of outflow. 
Therefore, multi-blackbody spectra are negligibly affected, either.
This has a profound implication when one performs spectral fitting of
black hole objects, notably of Ultraluminous X-ray Sources 
(ULXs, see subsection 4.3).

We here remark on the reasons why we stick to the one-dimensional model,
when multi-dimensional simulation data are available.
There are a number of reasons for this.
Although it has become possible to simulate the flow from the first principle,
they are still subject to numerical errors and limitations 
arising from the finite mesh spacing.
Also, it is not always easy to specify the physical processes from 
the vast simulation data with substantial fluctuations and numerical errors.
Further, 
multi-dimensional RHD simulations are very expensive and time-consuming.
That is, it is impossible to perform extensive parameter studies.
Therefore, studies based on simplified (one-dimensional) models
like the present one should be useful and beneficial 
for understanding the physics.

\subsection{Work exerted on outflow}
In our simplified one-dimensional model,
we consider loss of mass, angular momentum, and energy by outflow.
These effects are incorporated by spatial variation of the mass accretion rate.
I.e., we considered angular momentum and energy loss carried by outflow material,
assuming specific angular momentum and specific energy are the same for the outflow and the accretion flow.
There exists, however, another important factor which should be include in the energy equation;
that is the work exerted on the outflow by the disk material, $Q^{-}_{\rm wind}$.
Obviously, outflow cannot have positive energy to reach the infinity
without acquiring additional energy by the underlying accretion flow.
Poutanen et al. (2007), for example, include this effect
by assuming that radiative loss of the accreting material is partly
transferred to the outflow.

To see how our results are affected by this additional energy loss,
we reduce the specific energy of accretion flow by hand, 
assuming $Q^{-}_{\rm wind} = Q^{-}_{\rm rad}$.
The results are that the effective temperature profile hardly varies,
even though the amount of mass loss decreases.
Our conclusion of unchanged temperature profile is not altered.

\subsection{Model for Ultraluminous X-Ray Sources}
ULXs are bright, compact X-ray sources found in the off-center
region of nearby galaxies.
Their luminosities are typically,
$L_{\rm X} \sim 10^{39-41} [\rm erg/s]$,
and, hence, exceed the Eddington luminosity of a neutron star \citep{Fab89}.
There are two hypotheses for the origin of ULXs.
If the luminosity is below the Eddington luminosity,
it then follows that the mass of the central black holes of ULXs
 should be intermediate-mass black hole (IMBH), whose masses range over $10^{2-4} M_{\odot}$.
If the luminosity can be above the Eddington luminosity, conversely,
stellar-mass black holes can also account for the luminosities.
Unfortunately, the kinematic method, which is very powerful
to estimate the black hole masses in binary systems, is not applied
to the ULXs because no appreciable (optical) line features are observed.

Interestingly, ULXs share similar spectral features with the
black hole binaries (\cite{ColMus99}; \cite{Mak+00}; see review by \cite{Don+07a}).
Therefore, it seems possible to estimate the black hole mass through
the spectral fitting, although the results look controversial. 
Some authors claim that the ULXs should have IMBHs.
This is because the obtained blackbody temperature is low, 
on the order of $\sim$ 0.1 keV (\cite{Cro+04}; \cite{Mil+04}; \cite{Rob+05}).  
(The blackbody temperature is proportional to $M^{-1/4}$ for the same Eddington ratio.)
Some others claimed that the ULXs should contain stellar-mass black holes (\cite{Kin+01}; \cite{Wat+01}; \cite{Oka+06}; \cite{Vie+06}).
Vierdayanti et al. (2008), for example, claim that
the spectral fitting with the conventional spectral model
(with blackbody and power law spectral components)
is not reliable when the power-law component dominates, 
and demonstrated basing on the original slim-disk model that
some ULXs exhibit spectral signatures of the supercritical accretion flow
(see also \cite{Vie+06}).
The present study supports their conclusions, since, even if 
outflow effects are taken into account, neither of
the effective temperature profiles nor multi-color blackbody spectra are altered. 


\bigskip
We would like to thank K. Watarai and R. Kawabata for
useful comments and discussions. 
This work is supported in part by the Grant-in-Aid of MEXT (19340044, SM), and by the Grant-in-Aid for the global COE programs on gThe Next Generation of Physics, Spun from Diversity and Emergenceh from MEXT (SM),
Ministry of Education, Culture, Sports, Science, and
Technology (MEXT)
Young Scientist (B) 20740115 (KO).
Numerical computations were [in part] carried out on Cray XT4 at Center for Computational Astrophysics, CfCA, of National Astronomical Observatory of Japan.

\end{document}